\documentclass{elsart}

\usepackage{natbib}

\usepackage{amssymb}

\begin{document}

\begin{frontmatter}



\title{A first XMM-Newton look at the most X-ray-luminous galaxy cluster \\
RX J1347.5$-$1145}


\author{Myriam Gitti \& Sabine Schindler}

\address{Institut f\"ur Astrophysik, Leopold-Franzens Universit\"at Innsbruck,
Technikerstra\ss e 25, A-6020 Innsbruck, Austria}

\begin{abstract}
We present the first results from an \textit{XMM-Newton} observation of  
RX J1347.5$-$1145 (z=0.451), the most luminous X-ray cluster of galaxies 
currently known, with a luminosity $L_X = 6.0 \pm 0.1 \times 10^{45} 
\mbox{ erg s}^{-1}$  in the [2-10] keV energy band. 
The cluster has an overall temperature of $kT=10.0 \pm 0.3$ keV and is not
isothermal: the temperature profile shows a decline in the outer regions
and a drop in the centre, indicating the presence of a cooling core.
The spectral analysis identifies a hot region at 
radii 50-200 kpc to south-east of the main X-ray peak, at a position 
consistent with the subclump seen in the X-ray image. 
Excluding the data of the south-east quadrant, the cluster appears relatively
relaxed and we estimate a total mass within 1 Mpc of $1.0 \pm 0.2 
\times 10^{15} M_{\odot}$. 
\end{abstract}


\end{frontmatter}

% main text

%%%%%%%%%%%%%%%%%%%%%%%%%%%%%%%%%%%%%%%%%%%%%%%%%%%%%%%%%%%%%%%%%%%%%%%%%%%%%

\section{Introduction}

The cluster RX J1347.5$-$1145 detected in the \textit{ROSAT} All-Sky Survey
and further studied with \textit{ROSAT} HRI and \textit{ASCA} 
(Schindler et al. 1995, 1997) is exceptional in many aspects. It is the
most X-ray-luminous galaxy cluster known (Schindler et al. 1995), it
shows a very peaked X-ray emission profile and presents a strong cooling 
flow in its central region. Submm observations in its direction 
showed a very deep SZ decrement (Komatsu et al. 1999, 2001; 
Pointecouteau et al. 1999, 2001).
Due to the presence of gravitational arcs, this cluster is also well suited 
for a comparison of lensing mass and X-ray mass. 
Optical studies of weak lensing have been performed by Fischer \& Tyson (1997)
and Sahu et al. (1998).   
Recent \textit{Chandra} observations (Allen et al. 2002) discovered  a 
region of relatively hot, bright X-ray emission, located approximately 20 
arcsec to the south-east of the main X-ray peak at a position consistent
with the region of enhanced SZ effect. 

We present the first results from an \textit{XMM-Newton} observation of 
RX J1347.5$-$1145, which was performed in July 2002 during 
rev. 484. Standard processing is applied to prepare data and reject the soft 
proton flares. The exposure times after cleaning are 32.2 ks for MOS1,
32.5 ks for MOS2 and 27.9 ks for pn. The background estimates are obtained 
using a blank-sky observation consisting of several high-latitude pointings 
with sources removed (Lumb et al. 2002). The background subtraction 
(for spectra and surface brightness profiles) is performed as described in 
full detail in Arnaud et al. (2002).
The source and background events are corrected for vignetting using the
weighted method described in Arnaud et al. (2001).
RX J1347.5$-$1145 is at a redshift $z=0.451$. With a Hubble constant of
$H_0 = 70 \mbox{ km s}^{-1} \mbox{ Mpc}^{-1}$, and $\Omega_M = 
1-\Omega_{\Lambda} = 0.3$, the luminosity distance is 2506 Mpc and the angular
scale is 5.77 kpc per arcsec. 

%%%%%%%%%%%%%%%%%%%%%%%%%%%%%%%%%%%%%%%%%%%%%%%%%%%%%%%%%%%%%%%%%%%%%%%%%%%%%%

\section{Spectral analysis}
\label{spectral.sec}

\begin{figure}[ht]
\includegraphics{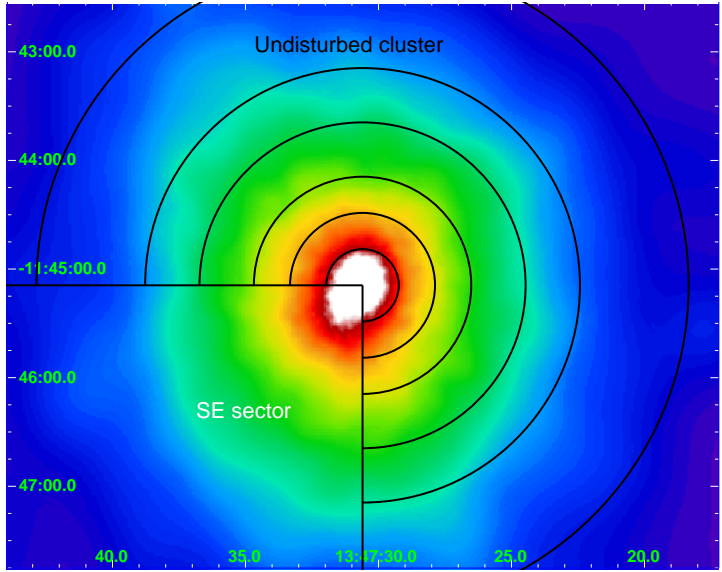}
\vspace{8.0cm}
\caption{
Total (MOS+pn) \textit{XMM--Newton} EPIC mosaic image of RX J1347.5$-$1145
in the [0.9-10] keV energy band. The image is corrected for vignetting and
exposure and is adaptively smoothed.}
\label{fig1}
\end{figure}
Since a morphological analysis indicates the presence of a substructure
$\sim$ 20 arcsec to the south-east (SE) of the X-ray peak
(already revealed in previous observations with \textit{Chandra}, Allen et al. 
2002), we perform the spectral analysis for the data excluding the SE quadrant
(hereafter undisturbed cluster) and for the full $360^{\circ}$ data, 
separately.
In both cases, the data are divided into the annular regions shown in Fig. 
\ref{fig1} and detailed in Table 1. A single spectrum is 
extracted for each region and then the data from the three cameras are 
modelled simultaneously using the XSPEC code, version 11.3.0. 
Spectral fitting is performed in the [0.5-8] keV band.
The spectra are modelled using a simple, single-temperature model
(MEKAL plasma emission code in XSPEC) with the absorbing column density 
fixed to the Galactic value ($N_{\rm H} = 4.85 \times 10^{20} 
\mbox{ cm}^{-2}$, Dickey \& Lockman 1990). The free parameters in this
model are the temperature $kT$, metallicity $Z$ (measured relative to the
solar values) and normalization (emission measure).
A typical simultaneous fit to the spectra extracted in each region is shown
in Fig. \ref{fig2_3}.
\begin{figure}
\includegraphics{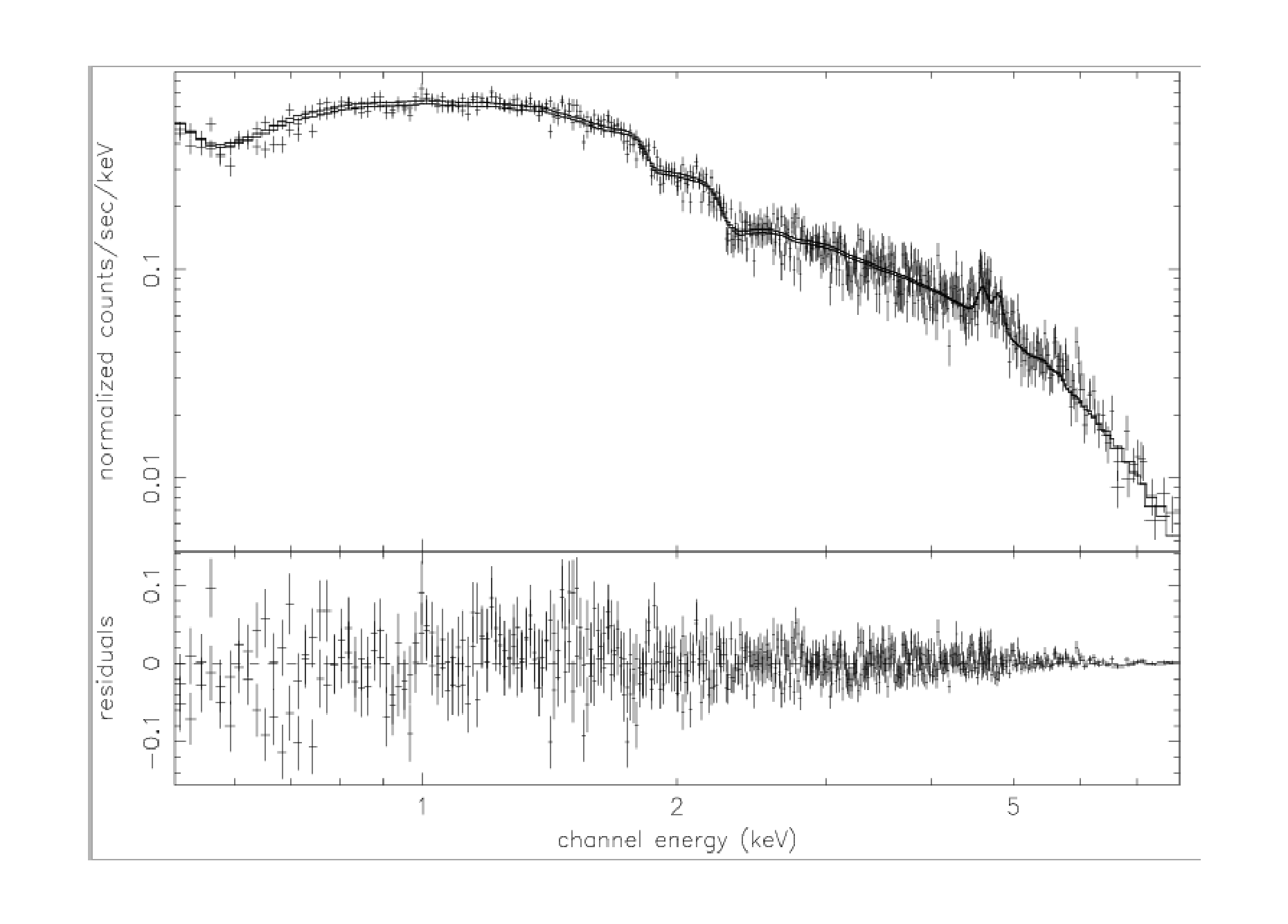}
\vspace{7.5cm}
%\caption{
%Global MOS spectrum integrated in a circular region of radius 5 arcmin. 
%The fit with a MEKAL model and the residuals are shown.
%}
%\label{fig2}
\end{figure}
\begin{figure}
\includegraphics{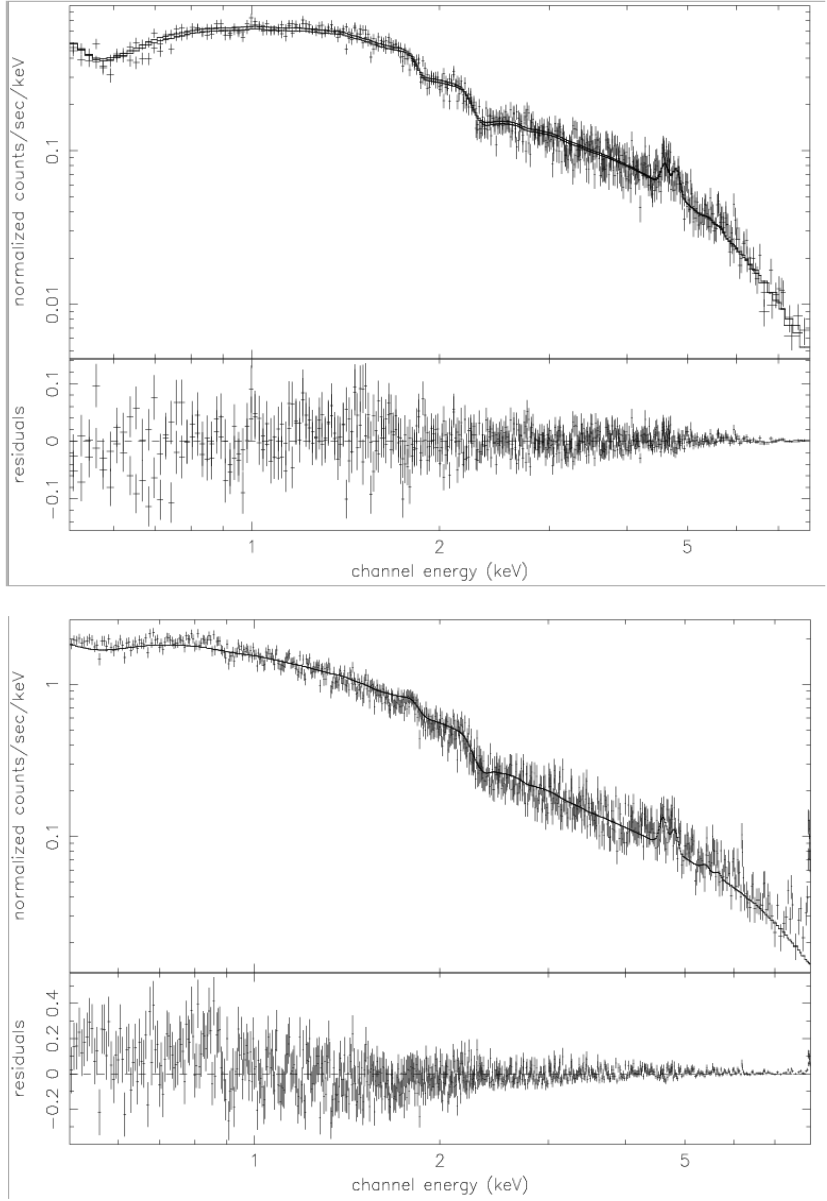}
\vspace{7.5cm}
\caption{
Global MOS (upper panel) and pn (lower panel) spectrum integrated in a 
circular region of radius 5 arcmin. 
The fit with a MEKAL model and the residuals are shown.
}
\label{fig2_3}
\end{figure}
The best-fitting parameter values and 90\% confidence levels derived from the
fits to the annular spectra are summarized in Table 1.
The projected temperature profiles determined with this model from the data
excluding the SE quadrant and for the full $360^{\circ}$ data are shown
in Fig. \ref{fig4}. 
The temperature of the undisturbed cluster rises from a mean value of
$8.9 \pm 0.3$ keV within 115 kpc to $kT = 11.1 \pm 1.0$ keV over the 0.1-0.5
Mpc region, then declines down to a mean value of $6.0^{+2.6}_{-1.6}$ keV 
in the outer regions (1.0-1.7 Mpc).
As a general trend, we note that the temperature of the undisturbed cluster is 
lower than that estimated from the full $360^{\circ}$ data, even though 
consistent within the errors.
The derived temperature profile indicates the existence of a cool core: 
even though the gas temperature in the central 100 kpc is hotter than that 
usually found in cooling core clusters, there is probably a much cooler 
gas in the very center, which cannot be resolved due to the relatively 
high cluster redshift. 
\begin{table}
\small
\hskip 0.2truein
\caption
{
The results from the spectral fitting in concentric annular regions for the
undisturbed cluster and the full $360^{\circ}$ data. 
Temperatures ($kT$) are in keV and metallicities ($Z$)
in solar units. The total $\chi^2$ values and numbers of degrees of freedom 
(DOF) in the fits are listed in columns 4 and 7. Errors are 90\% 
confidence levels. 
}
\vskip 0.1truein
\hskip 0.0truein
\begin{tabular}{|c|ccc|ccc|}
\hline
~ & \multicolumn{3}{|c|}{SE sector excluded} & \multicolumn{3}{|c|}{full $360^{\circ}$} \\
\hline  
Radius (kpc)   &     $kT$    & $Z$    &     $\chi^2$/DOF  & $kT$    & $Z$    &     $\chi^2$/DOF   \\
\hline
0-115 & $8.9^{+0.3}_{-0.3}$ & $0.34^{+0.05}_{-0.05}$ & 982/880 & $9.2^{+0.3}_{-0.3}$ & $0.31^{+0.04}_{-0.04}$ & 1183/1026 \\
115-230 & $10.7^{+0.7}_{-0.6}$ & $0.26^{+0.08}_{-0.08}$ & 696/664 & $11.4^{+0.6}_{-0.5}$ & $0.25^{+0.07}_{-0.07}$ & 970/855\\
230-345 & $11.9^{+1.6}_{-1.3}$ & $0.16^{+0.14}_{-0.15}$ & 433/384 & $11.8^{+1.0}_{-0.8}$ & $0.24^{+0.10}_{-0.10}$ & 610/564\\
345-520 & $10.7^{+1.1}_{-1.0}$ & $0.24^{+0.13}_{-0.13}$ & 350/341 & $11.4^{+1.1}_{-0.9}$ & $0.27^{+0.11}_{-0.11}$ & 471/493\\
520-690 & $9.0^{+1.4}_{-1.1}$ & $0.16^{+0.18}_{-0.16}$ & 239/210 & $9.4^{+1.3}_{-1.0}$ & $0.21^{+0.16}_{-0.16}$ & 372/300\\
690-1040 & $9.4^{+2.1}_{-1.4}$ & $0.19^{+0.26}_{-0.19}$ & 293/264 & $9.0^{+1.5}_{-1.2}$ & $0.27^{+0.20}_{-0.23}$ & 436/346\\
1040-1730 & $6.0^{+2.6}_{-1.7}$ & $0.40^{+0.50}_{-0.37}$ & 593/421 & $7.4^{+3.6}_{-1.6}$ & $0.30^{+0.42}_{-0.30}$ & 977/542\\
0-1730 & $9.4^{+0.3}_{-0.3}$ & $0.26^{+0.04}_{-0.04}$ & 1957/1452 & $10.0^{+0.3}_{-0.2}$ & $0.26^{+0.03}_{-0.03}$ & 2586/1679\\
\hline                                   
\end{tabular}
\label{profilot.tab}
 \end{table}
\begin{figure}[ht]
\includegraphics{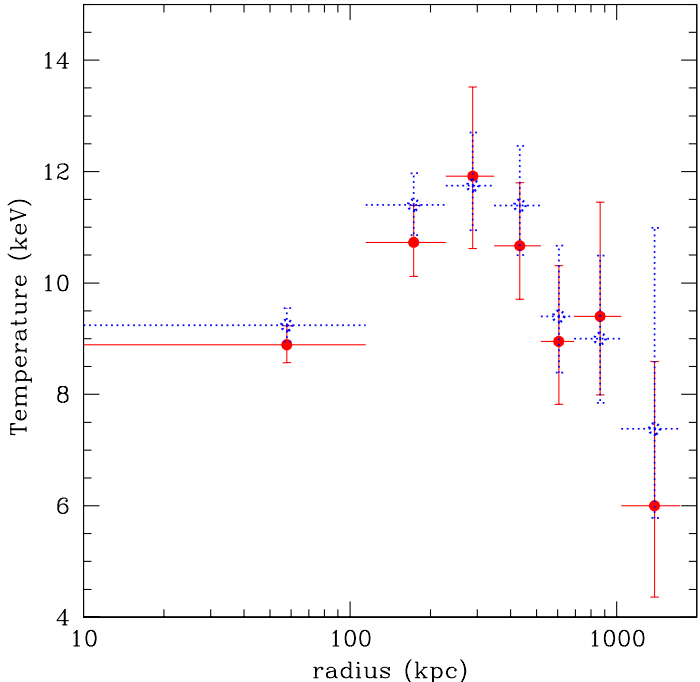}
\vspace{9.2cm}
\caption{
\textit{Full circles}: the projected X-ray gas temperature profile measured 
from data excluding the SE quadrant (undisturbed cluster).
\textit{Dotted circles}: same as full circles, but from the full 
$360^{\circ}$ data.}
\label{fig4}
\end{figure}
The metallicity profile is consistent with being constant, with an overall
value of $Z=0.26 \pm 0.04 Z_{\odot}$.
Within the radius of 1.7 Mpc, a fit to the full 360$^{\circ}$ data
gives an overall $kT = 10.0 \pm 0.3$ keV, $Z=0.26 \pm 0.03 Z_{\odot}$ and 
$L_X \mbox{ (2-10 keV)}= 6.0 \pm 0.1 \times 10^{45} \mbox{ erg s}^{-1}$. 

In order to study in more detail the thermal structure of the cluster, we
extract the spectrum in the annular region including the substructure
identified by the morphological analysis and compare the results in different
directions. A fit to the data for the SE quadrant between
radii 50-200 kpc yields a best-fitting temperature $kT= 12.7 ^{+1.0}_{-0.9}$ 
keV, while in other directions the mean value is $kT=10.2 \pm 0.5$ keV. 
Therefore, the spectral analysis confirms that the region 
corresponding to the subclump seen in the X-ray image
is significantly hotter than the surrounding gas.

%%%%%%%%%%%%%%%%%%%%%%%%%%%%%%%%%%%%%%%%%%%%%%%%%%%%%%%%%%%%%%%%%%%%%%%%%%%%%%

\section{Spatial analysis and mass determination}

\begin{figure}
\includegraphics{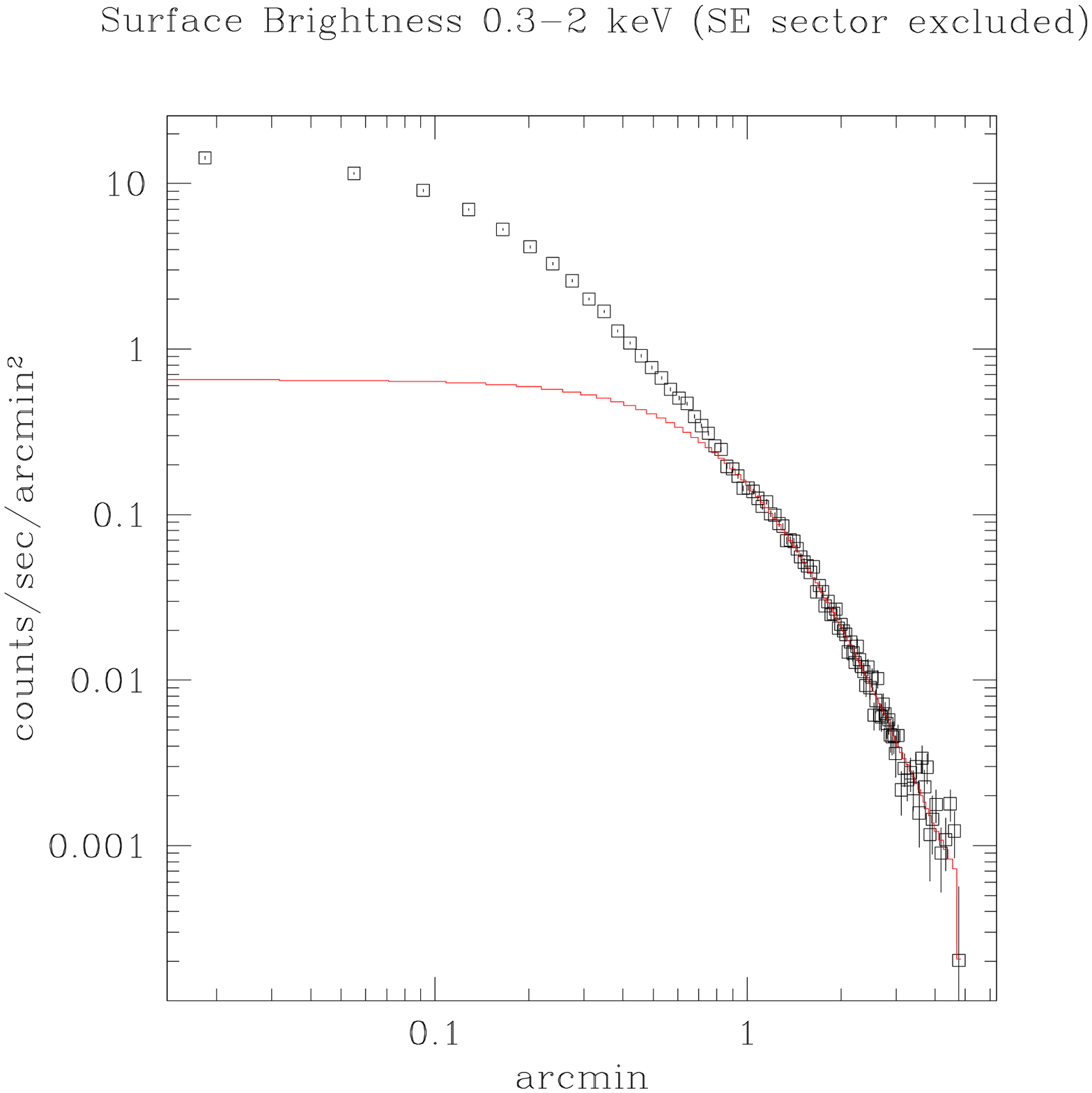}
\vspace{9.0cm}
\caption{
Background subtracted, azimuthally averaged radial surface brightness profile 
for the data excluding the SE quadrant in the [0.3-2] keV range. 
The line shows the $\beta$-model fit to the outer regions. A clear excess
in the center is visible.}
\label{fig5}
\end{figure}

We compute a background-subtracted vignetting-corrected radial surface
brightness profile for the undisturbed cluster in the [0.3-2] keV energy band. 
The surface brightness profile is then fitted with a $\beta$-model 
(Cavaliere \& Fusco Femiano 1976). We find that for 350 kpc - 1.7 Mpc the data 
can be described by a $\beta$-model with a core radius $r_{\rm c}=367 \pm 3$ 
kpc and a slope parameter $\beta=0.93 \pm 0.01$. However, a single 
$\beta$-model is not a good description of the entire profile, as the fit to 
the outer regions shows a strong excess in the centre when compared to the
model (see Fig. \ref{fig5}).
The peaked emission is a strong indication for a cooling core in this 
cluster. 

The total gravitating mass distribution is calculated under the usual 
assumptions of hydrostatic equilibrium and spherical symmetry 
using the deprojected density distribution calculated from
the parameters of the $\beta$-model.
Only data beyond $30''$ ($\sim 175$ kpc) are considered: in the central
bins the temperature as estimated in Sect. \ref{spectral.sec} is 
affected by the \textit{XMM} PSF and projection effects, while for the
outer regions these effects can be neglected (e.g. Kaastra et al. 2004). 
Within 1 Mpc we find a total mass of $1.0 \pm 0.2 \times 10^{15} M_{\odot}$,
while within 1.7 Mpc the total mass is $2.0 \pm 0.4 \times 10^{15} M_{\odot}$. 
These results are in agreement with \textit{Chandra} (Allen et al. 2002) 
and weak lensing analysis (Fischer \& Tyson 1997) results and
slightly higher than that derived by \textit{ROSAT/ASCA} 
(Schindler et al. 1997). 

%%%%%%%%%%%%%%%%%%%%%%%%%%%%%%%%%%%%%%%%%%%%%%%%%%%%%%%%%%%%%%%%%%%%%%%%%%%%

\section{Conclusions}

The \textit{XMM-Newton} observation of RX J1347.5$-$1145 confirms that it 
is, with a luminosity $L_X = 6.0 \pm 0.1 
\times 10^{45} \mbox{ erg s}^{-1}$ (2-10 keV energy band), 
the most X-ray-luminous cluster discovered to date. RX J1347.5$-$1145 
is a hot cluster (overall temperature: $kT=10.0 \pm 0.3$ keV),
not isothermal: the temperature profile shows the presence of 
a cool core and a decline of the temperature in the outer regions. 
The spectral analysis identifies a relatively
hot region at radii 50-200 kpc to the SE of the main X-ray peak. 
This hot region is found at the same position as the subclump seen in 
the X-ray image.
On the other hand, excluding the data of the SE quadrant
the cluster appears relatively relaxed and we estimate a total mass within 
1 Mpc of $1.0 \pm 0.2 \times 10^{15} M_{\odot}$.

%%%%%%%%%%%%%%%%%%%%%%%%%%%%%%%%%%%%%%%%%%%%%%%%%%%%%%%%%%%%%%%%%%%%%%%%%%

\section*{\textit{Acknowledgements}}

This work is based on observations obtained with \textit{XMM-Newton}, 
an ESA science mission with instruments and contributions directly funded 
by ESA Member States and the USA (NASA).
M.G. would like to thank E. Belsole, A. Castillo-Morales, S. Majerowicz 
and E. Pointecouteau for suggestions concerning \textit{XMM-Newton} 
data analysis, and S.Ettori for his advices in the spectral analysis.
This work was supported by the Austrian Science Foundation FWF under 
grant P15868, {\"O}AD Amad{\'e}e Projekt 18/2003 and
{\"O}AD Acciones Integradas Projekt 22/2003.

%%%%%%%%%%%%%%%%%%%%%%%%%%%%%%%%%%%%%%%%%%%%%%%%%%%%%%%%%%%%%%%%%%%%%%%%%%%%

% \appendix

% \section{}
% \label{}

% Bibliographic references with the natbib package:
% Parenthetical: \citep{Bai92} produces (Bailyn 1992).
% Textual: \citet{Bai95} produces Bailyn et al. (1995).
% An affix and part of a reference:
%   \citep[e.g.][Ch. 2]{Bar76}
%   produces (e.g. Barnes et al. 1976, Ch. 2).

\end{document}